\documentstyle [12pt] {article}
\begin{document}

\centerline{\bf PION NUCLEON COUPLING CONSTANT, GOLDBERGER-TREIMAN}
\centerline{\bf DISCREPANCY AND $\pi N$ $\sigma$ TERM}
\vskip 0.5cm
\centerline{Miroslav Nagy\footnote{E-mail address: fyzinami@savba.sk}}
\centerline{\sl Institute of Physics, Slovak Academy of Sciences, 
845 11 Bratislava, Slovakia}
\centerline{Michael D. Scadron\footnote{E-mail address:
scadron@physics.arizona.edu}}
\centerline{\sl Physics Department, University of Arizona, Tucson,
Arizona 85721,USA}
\centerline{Gerald E. Hite\footnote{E-mail address:
hiteg@tamug.edu}}
\centerline{\sl Texas A\&M University at Galveston, TX 77553, USA}

\vskip 1.0cm
\begin{abstract}
We start by studying the Goldberger-Treiman discrepancy (GTd) $\Delta =
(2.259\pm 0.591)\%$. Then we look at the $\pi N$ $\sigma$ term, with the
dimensionless ratio $\sigma_N/2m_N=3.35\%$. Finally we return to
predicting (via the quark model) the $\pi N$ coupling constant, with
GTd $\Delta\to 0$ as $m_q\to m_N/3$.
\end{abstract}
\vskip1.0truecm
\noindent Given the recent new value of the $\pi NN$ coupling constant
\cite{Bugg}
\begin{equation}
g^2_{\pi NN}/4\pi = 13.80\pm 0.12 ~~~{\rm or}~~~ g_{\pi NN}=13.169\pm
0.057,  \label{g^2}
\end{equation}
along with the observed axial current coupling \cite{PDG}
\begin{equation}
g_A = 1.267 \pm 0.004, \label{g_A}
\end{equation}
combined with the measured pion decay constant \cite{PDG}
\begin{equation}
f_{\pi} = (92.42\pm 0.26) {\rm MeV}, \label{f_pi}
\end{equation}
the Goldberger-Treiman discrepancy (GTd) is then
\begin{equation}
\Delta = 1 - \frac{m_Ng_A}{f_{\pi}g_{\pi NN}} = (2.259\pm 0.591)\% .
\label{Delta}
\end{equation}
Here we have used the mean nucleon mass $m_N$= 938.9 MeV and have
computed the overall mean square error.

To verify this GTd in Eq.(\ref{Delta}), we employ the constituent quark
loop with imaginary part \cite{Coon}
\begin{equation}
{\rm Im} f_{\pi}(q^2) = \frac{3g_{\pi qq}}{2}\frac{4{\hat
m}}{8\pi}\left(1 -\frac{4{\hat m}^2}{q^2} \right)^{1/2}\Theta(q^2 -
4{\hat m}^2) .  \label{Imag}
\end{equation}
This follows from unitarity with the inclusion of a factor of 3 from
colour. Following ref. \cite{Coon} using the quark level
Goldberger-Treiman relation $f_{\pi}g_{\pi qq} = {\hat m}$, the GTd to
fourth order in $q'^2$ predicts for a once-subtracted dispersion
relation assuming a quark-level GTR:
\begin{equation}
\frac{f_{\pi}(q^2) - f_{\pi}(0)}{f_{\pi}(0)} = \frac{q^2}{\pi}
\int_{4{\hat m}^2}^{\infty}\frac{dq'^2\left(1-\frac{4{\hat m}^2}{q'^2}
\right)^{1/2}}{q'^2(q'^2-q^2)}\frac{3g^2_{\pi qq}}{4\pi} , \label{frac}
\end{equation}
or for $q^2=m_{\pi}^2$, the integral in Eq.(\ref{frac}) gives a
discrepancy for $f_{\pi}$
\begin{equation}
{\bar \Delta} = \frac{f_{\pi}(m^2_{\pi})}{f_{\pi}(0)} - 1 =
\frac{3g^2_{\pi qq}}{2\pi^2}\left[1-r{\rm tan}^{-1}\left(\frac{1}{r}
\right)\right]  \label{Delta1}
\end{equation}
for $r^2=\frac{4{\hat m}^2}{m^2_{\pi}}-1\geq 0$. Since $m^2_{\pi}/
4{\hat m}^2\ll 1$, a Taylor series expansion leads to
\[
1-r{\rm tan}^{-1}\left(\frac{1}{r}\right) = \frac{1}{3r^2}-
\frac{1}{5r^4}+...= \frac{m^2_{\pi}}{12{\hat m}^2}\left(1+
\frac{1}{10}\frac{m^2_{\pi}}{{\hat m}^2}+...\right)
\]
and a discrepancy \footnote{From Dwight Integral tables, 
Eq.(\ref{Delta1}) above stems from Eq.122.1 on p.31, and the needed
Taylor series of Eq.505.1, p.118: ${\rm tan}^{-1}x=x-\frac{x^3}{3}
+\frac{x^5}{5}+...,$ for $x<1$.}
\begin{equation}
{\bar \Delta} = \frac{f_{\pi}(m^2_{\pi})}{f_{\pi}(0)} - 1 =
\frac{m^2_{\pi}}{8\pi^2f_{\pi}^2}\left(1+\frac{1}{10}
\frac{m^2_{\pi}}{{\hat m}^2}\right) \approx 2.946\% . 
\label{Delta'}
\end{equation}
The first term on the rhs is independent of ${\hat m}$, while in the
small second term we take ${\hat m}= m_N/3$. This then leads to a
net 2.946\% correction in Eq.(\ref{Delta'}).

Since the physical GT relation becomes exact $(f_{\pi}g_{\pi NN}=
m_Ng_A)$ when $m_{\pi}\to 0$ for a conserved axial current, it should
not be surprising that the measured GTd in Eq.(\ref{Delta}) of $(2.259
\pm 0.591)\%$ is within 1.16 standard deviations from the
dispersion-theoretical $\bar{\rm GTd}$ ${\bar \Delta} = 2.946 \%$ in
Eq.(\ref{Delta'}). Appreciate that $g_A$ is measured at $q^2 = 0$ while
$f_{\pi}$ is measured at $q^2 = m^2_{\pi}$ but $f_{\pi}(0)$ is inferred
at $q^2=0$ via Eq.(\ref{Delta'}).

Just as the chiral-breaking $SU(2)$ GTd is 2--3\%, the
$SU(2)\times SU(2)$ $\pi N$ $\sigma$ term of 63 MeV corresponds to a
dimensionless ratio of about 3\%:
\begin{equation}
\frac{\sigma_N}{2m_N}=\frac{63~{\rm MeV}}{2\times 938.9~ {\rm MeV}}
\approx 3.35\% .   \label{sigma}
\end{equation}
Alternatively the chiral-limiting (CL) nucleon mass is related to the 
$\pi N$ $\sigma$ term as \cite{GMcNS}
\begin{equation}
m_N^2 = (m^{CL}_N)^2 +m_N\sigma_N,~~~ {\rm or~ with}~~
\sigma_N=63~{\rm MeV},
\label{mnmCL}
\end{equation}
\begin{equation}
\frac{m_N}{m_N^{CL}} - 1 = 3.53\%,~~~ {\rm with}~~ 
m_N^{CL}=906.85~{\rm MeV}. \label{fracmn}
\end{equation}
Note the many $3\%$ CL relations in Eqs.
(\ref{Delta}),(\ref{Delta'}),(\ref{sigma}),(\ref{fracmn}) above. Now we 
justify the $\sigma$ term $\sigma_N=63$ MeV.

The explicit $SU(2)\times SU(2)$ chiral-breaking $\sigma$ term is the
sum of the perturbative GMOR \cite{GMOR} or quenched APE \cite{APE} part
\begin{equation}
\sigma_N^{GMOR} = (m_{\Xi} + m_{\Sigma} - 2m_N) \frac{m^2_{\pi}}{m^2_K
- m^2_{\pi}} = 26~ {\rm MeV},  \label{GMOR}
\end{equation}
\begin{equation}
\sigma_N^{APE} = (24.5 \pm 2)~ {\rm MeV},  \label{APE}
\end{equation}
plus the nonperturbative linear $\sigma$ model (L$\sigma$M) nonquenched
part \cite{GML} due to $\sigma$ tadpoles for the chiral-broken
$m^2_{\pi}$ and $\sigma_N$, with ratio predicting
\begin{equation}
\sigma_N^{L\sigma M} = \left(\frac{m_{\pi}}{m_{\sigma}}\right)^2 m_N
\approx 40~ {\rm MeV}   \label{LsM}
\end{equation}
for $m_{\sigma}\approx 665$ MeV \cite{SKN}, a model-independent coupled
channel dispersion theory and parameter-free relation. Specifically,
Eq.(\ref{LsM}) stems
from semi-strong L$\sigma$M tadpole graphs generating
$\sigma_N$ and $m^2_{\pi}$. Their ratio cancels out the 
$\langle\sigma|H_{ss}|0\rangle$ factor. The L$\sigma$M couplings
$2g_{\sigma\pi\pi}$=$m_{\sigma}^2/f_{\pi}$ and $f_{\pi}g_{\sigma
NN}=m_N$ then give $\sigma^{L\sigma M}_N$ = $(m_{\pi}/m_{\sigma})^2m_N$
as found in Eq.(\ref{LsM}).
Since the $\sigma(600)$ has been observed
\cite{PDG}, with a broad width, but the central model-independent value 
\cite{SKN} is known to be 665 MeV, the chiral L$\sigma$M mass ratio in 
Eq.(\ref{LsM}) is expected to be quite accurate - while being free of 
model-dependent parameters. The authors of \cite{TaR} find the
$\sigma$ meson between 400 MeV and 900 MeV, with the average mass 650
MeV near 665 MeV from \cite{SKN}. Then the sum of (\ref{GMOR},\ref{APE}) 
plus (\ref{LsM}) is
\begin{equation}
\sigma_N = \sigma_N^{GMOR,APE} + \sigma_N^{L\sigma M}\approx (25+40) 
~{\rm MeV}= 65 ~{\rm MeV}.   \label{sigmaN}
\end{equation}
Rather than add the perturbative plus nonperturbative parts as in
Eq.(\ref{sigmaN}), one can instead work in the infinite momentum frame
(IMF) requiring squared masses \cite{MDS} and only one term (tadpole
terms ${}\to 0$ in the IMF) \cite{CSS}
\begin{equation}
\sigma_N^{IMF} = \frac{m^2_{\Xi}+m^2_{\Sigma}-2m^2_N}{2m_N}\left(
\frac{m^2_{\pi}}{m^2_K-m^2_{\pi}}\right) = 63~ {\rm MeV}.  \label{IMF}
\end{equation}
Note that Eqs.(\ref{sigmaN}) and (\ref{IMF}) are both very near the
observed value \cite{Koch} $(65\pm 5)$ MeV.

With hindsight, we can also deduce the $\pi N$ $\sigma$ term via 
PCAC (partially conserved axial current) at the Cheng-Dashen (CD)
point \cite{ChDa} with background isospin-even $\pi$N amplitude
\begin{equation}
{\bar F}^+(\nu=0,t=2m^2_{\pi})= \sigma_N/f^2_{\pi} + O(m^4_{\pi}).
\label{F+nu}
\end{equation}
At this CD point, a recent Karlsruhe data analysis by G. H\"ohler
\cite{Koch} finds
\begin{equation}
{\bar F}^+(0,2m^2_{\pi})= \sigma_N/f^2_{\pi} + 0.002m^{-1}_{\pi}=
1.02m^{-1}_{\pi},  \label{F2mpi}
\end{equation}
implying $\sigma_N=63$ MeV for $f_{\pi}=93$ MeV, $m_{\pi}=139.57$
MeV.

We can unify the earlier parts of this paper by first inferring
from Eq.(\ref{Delta'}) the chiral limit (CL) pion decay constant 
\begin{equation}
f^{CL}_{\pi} = f_{\pi}/1.029 46\approx 89.775~{\rm MeV}  \label{fpiCL}
\end{equation}
using Eq.(\ref{Delta'}) and the observed \cite{PDG} $f_{\pi}=(92.42\pm
0.26)$ MeV. Then the quark-level GTr using the meson-quark coupling
$g=2\pi/\sqrt 3$ \cite{Elias} predicts the nonstrange quark mass in the
CL as
\begin{equation}
{\hat m}^{CL} = f_{\pi}^{CL}g = 325.67 ~{\rm MeV},  \label{mCL}
\end{equation}
close to the expected ${\hat m}^{CL} = m_N/3\approx 313$ MeV. This in
turn predicts the scalar $\sigma$ mass in the CL as \cite{GML,NJL}
\begin{equation}
m^{CL}_{\sigma}= 2{\hat m}^{CL} = 651.34~ {\rm MeV}   \label{msCL}
\end{equation}
and then the on-shell L$\sigma$M $\sigma$ mass obeys
\begin{equation}
m^2_{\sigma} - m^2_{\pi} = (m^{CL}_{\sigma})^2\approx (651.34~{\rm 
MeV})^2 ~~~{\rm or}~~~ m_{\sigma}\approx 665.76~{\rm MeV},  \label{ms2}
\end{equation}
almost exactly the model-independent $\sigma$ mass found in ref.
\cite{SKN}, also predicting $\sigma_N^{L\sigma M}$ in Eq.(\ref{LsM}).

In this letter we have linked the GT discrepancy
Eqs.(\ref{Delta}),(\ref{Delta'}) and the $\pi N$ $\sigma$ term
Eqs.(\ref{sigmaN}),(\ref{IMF}) with the L$\sigma$M values
Eqs.(\ref{fpiCL})-(\ref{ms2}). The predicted L$\sigma$M value of $g_{\pi
NN}$ is
\begin{equation}
g_{\pi NN} = N_c gg_A= 3(2\pi/\sqrt3)1.267\approx 13.79,  \label{gpiNN}
\end{equation}
near the observed value in Eq.(\ref{g^2}) with meson-quark coupling g.
Substituting Eq.(\ref{gpiNN}) into the GTd (Eq.(\ref{Delta})) in turn
predicts in the quark model
\begin{equation}
\Delta = 1 - \frac{m_N}{3m_q}\to 0~~~{\rm as}~~~ m_q\to m_N/3.
\label{DmN}
\end{equation}
However meson-baryon couplings for pseudoscalars (P), axial-vectors 
(A) and $SU(6)$-symmetric states are known \cite{MMN} to obey
\begin{equation}
(d/f)_P\approx 2.0, ~~~ (d/f)_A\approx 1.74, ~~~ (d/f)_{SU(6)}
= 1.50, \label{SU6}
\end{equation}
where the scales of $d,f$ characterize the symmetric, antisymmetric
$SU(3)$ structure constants. Note that the ratio remains the same:
\begin{equation}
\frac{(d/f)_A}{(d/f)_P}=\frac{1.74}{2.0}=0.87,~~~
\frac{(d/f)_{SU(6)}}{(d/f)_A}=\frac{1.50}{1.74}\approx 0.86.  
\label{dfAP}
\end{equation}
Thus to predict the quark-based $\pi NN$ coupling constant we weight
Eq.(\ref{gpiNN}) by the scale factor of Eq.(\ref{dfAP}) in order to
account for the $SU(6)$ quark content of $g_A$:
\begin{equation}
g_{\pi NN}= 3\times 2\pi/{\sqrt 3}\times 1.267\times 0.87\approx 12.00
\label{gpiNN'}
\end{equation}
and this predicted coupling constant is near 13.169 from ref.
\cite{Bugg}, or 13.145 from ref. \cite{AWP}, or nearer still to
13.054 from ref. \cite{STS}. One could alter this 0.87 reduction of
$g_A$ in Eq.(\ref{gpiNN'}) by using the quark-based factor 3/5=0.6,
where the $SU(6)$ factor for $g_A$ of 5/3 becomes inverted for
quarks as suggested in \cite{TNP}. In any case the predicted 
$\pi NN$ coupling lies between 12.00 and 13.79 in 
Eqs.(\ref{gpiNN'}),(\ref{gpiNN}), midway near the recent data in 
Eq.(\ref{g^2}).

In passing, we note that the large model-independent \cite{SKN} scalar
$\sigma$ mass of $m_{\sigma}\approx 665$ MeV is recovered via the
L$\sigma$M combined with the CL quark-level GTR
Eqs.(\ref{fpiCL})-(\ref{ms2}). Also the large almost
model-independent interior dispersion relation version of the $\pi N~
\sigma$ term \cite{HJS,KH} is between 65-80 MeV. While this $\sigma$
term follows from the two GMOR + L$\sigma$M terms in Eq.(\ref{sigmaN})
or from the IMF term in Eq.(\ref{IMF}), original chiral perturbation
theory (ChPT) of the 1970s suggested \cite{HP} $\sigma_N\approx 25$ MeV
near the GMOR value.

Modern ChPT now predicts \cite{GLS} a $\sigma_N$ of 45 MeV at $t=0$
extended up to the above presumably measured value of 60 MeV according
to \cite{GLS,HL}
\begin{equation}
60~{\rm MeV} = \sigma^{\rm GMOR}_N(25~{\rm MeV}) + \sigma_N^{\rm
higher~order~ChPT}(10~{\rm MeV}) + \sigma_N^{\rm t-dep.}(15~{\rm MeV})
+ \sigma_N^{{\bar s}s}(10~{\rm MeV})   \label{60MeV}
\end{equation}
and the latter ''three pieces happen to have the same sign as
$\sigma_N^{\rm GMOR}$'' \cite{HL}.

In summary, as $m_{\pi}\to 0$, $\partial A_{\pi}\to 0$, the quark-level
GT relation requires the observed $2-3\%$ GTd and $3\%$ $\sigma$ term
ratio to predict $g_{\pi NN}$, with $\Delta\to 0$ as $m_q\to m_N/3$ or
${\bar \Delta}\to 0$ when $m^2_{\pi}\to 0$. We have computed the $\pi N$
$\sigma$ term in many. different ways to find approximately $\sigma_N
=63$ MeV.

\noindent Acknowledgements: One of us (MDS) appreciates D.V. Bugg's
help in first obtaining Eq.(\ref{Delta'}) via a computer. Nonetheless
the reader may be more convinced with Eq.(\ref{Delta'}) using the
Dwight integral solution combined with the Taylor series in footnote 4.
This work is in part supported by the Slovak Agency for Science, Grant
2/3105/23.

\eject
\end{document}